\documentstyle[aps,prd,epsf,floats]{revtex}

\newcommand{\bea}{\begin{eqnarray}}
\newcommand{\eea}{\end{eqnarray}}
\newcommand{\bra}[1]{\langle{#1}|}
\newcommand{\ket}[1]{|{#1}\rangle}

\begin{document}

\preprint{astro-ph/0106197}
\draft
\twocolumn[\hsize\textwidth\columnwidth\hsize\csname
@twocolumnfalse\endcsname


\title{Evolution of cosmological perturbations
       in non-singular string cosmologies}

\author{Cyril Cartier${}^{(a)}$ and Jai-chan Hwang${}^{(b,c)}$
        and Edmund J. Copeland${}^{(a)}$}

\address{$^{(a)}$
         Centre for Theoretical Physics, University of Sussex,
         Falmer, Brighton BN1 9QJ, UK}

\address{$^{(b)}$
         Department of Astronomy and Atmospheric Sciences,
         Kyungpook National University, Taegu, Korea}

\address{$^{(c)}$
         Institute of Astronomy, Madingley Road, Cambridge, UK}

\date{\today}

\maketitle


\begin{abstract}
In a class of non-singular cosmologies derived from higher-order
corrections to the low-energy bosonic string action, we derive
evolution equations for the most general cosmological scalar,
vector and tensor perturbations. In the large scale limit, the
evolutions of both scalar and tensor perturbations are
characterised by conserved quantities, the usual curvature
perturbation in the uniform-field gauge and the tensor-type
perturbed metric. The vector perturbation is not affected, being
described by the conservation of the angular momentum of the fluid
component in the absence of any additional dissipative process.
{}For the scalar- and tensor-type perturbations, we show how,
given a background evolution during kinetic driven inflation of
the dilaton field, we can obtain the final power spectra generated
from the vacuum quantum fluctuations of the metric and the dilaton
field during the inflation.
\end{abstract}

\pacs{PACS numbers: 04.30.-w, 04.60.-m, 98.80.-k, 98.80.Hw
      \hfill  SUSX-TH-01-026}

\vskip2pc]


\section{Introduction}

Recent observations \cite{CMBR} of anisotropies in the cosmic
microwave background radiation strongly support the paradigm of
inflation as being the source of the primordial density
fluctuations, producing an almost flat power spectrum
\cite{Inflation}. The usual inflation models involve an
accelerated period being driven by the energy density associated
with the inflaton potential. However, there exist a class of
models which lead to inflation simply because they are driven by
the kinetic energy of a scalar field. There are a large class of
such pole-like inflation models: those derived from induced
gravity \cite{I-grav}, scalar-tensor gravity \cite{SC-grav} or the
pre-Big Bang scenario (PBB) \cite{PBB} of string theory.
Unfortunately, most of these models face the problem that they
tend to yield blue spectra for both the scalar- and tensor-type
perturbations with spectral indices $n_S \simeq 4$ and $n_T \simeq
3$, respectively, whereas the observationally supported Zel'dovich
spectra correspond to $n_S \simeq 1$ and $n_T \simeq 0$
\cite{Brustein:1995kn,Hwang:1998fp,Copeland:1998ie,Lidsey:1999mc}.
A possible resolution of this problem for the scalar perturbations
was proposed in \cite{Copeland:1997ug}, where fluctuations of the
axion field present in the low-energy string action can generate
the observed spectral index. However, it remains the case that the
tensor spectrum is difficult to reconcile with this class of
kinetic driven inflation models.

Perhaps the biggest issue facing gravity models is how to tackle
the initial curvature singularity. In the context of PBB, an
interesting suggestion was made that the graceful exit problem
could be resolved by including the quantum backreaction effect
\cite{backreaction}. A different approach uses an expansion in
terms of the inverse string tension $(\alpha')$ and coupling
corrections. This has already met with some success
\cite{pbb-higher-corr,Cartier:1999vk}. Given that there exist a
class of non-singular cosmologies based on these higher-order
corrections, it is natural to investigate the effect of these
correction terms on the evolution of primordial fluctuations. In
fact the impact of potential higher-order curvature corrections on
the evolution of the perturbations has recently been studied in
various inflationary models derived from string theory
\cite{Maggiore:1997bg,Kawai}. In particular, the classical
evolution in the presence of a Gauss-Bonnet coupling term was
considered in \cite{Gasperini:1997up,Hwang:1999gf}. Effects of a
string theory motivated axion coupling term $g(\phi) R \tilde R$
in the Lagrangian, where $R \tilde R \equiv
\eta^{\mu\nu\sigma\tau} R_{\mu\nu}^{\;\;\;\;\epsilon\kappa}
R_{\sigma\tau\epsilon\kappa}$, was investigated in
\cite{Choi:1999zy}. The general feature that appears to emerge is
that the higher-order curvature corrections generally flatten the
spectral distributions of primordial vacuum fluctuations. Hence,
one may imagine a scenario where the observationally relevant
perturbations for the large scale structures leave the Hubble
radius during a highly-curved regime, leaving the power-spectrum
nearly scale-invariant.

The aim of this paper is to calculate the vacuum fluctuations
arising out of the most general non-singular solutions formed to
date from the first order curvature corrections, corrections which
could arise in the context of the massless bosonic sector of the
low-energy effective action of string theory.

The paper is organised as follows. In \S \ref{sec:Action} we
introduce a general action including possible curvature
corrections up to fourth order in derivatives and we derive the
corresponding field equations. \S \ref{sec:Perturbation} is
devoted to studying the classical evolution of three types of
perturbations. We derive closed form equations for both scalar-
and tensor-type perturbations and determine the corresponding
large-scale exact solutions. We also show that the vector-type
perturbations are described by the conservation of the angular
momentum of the additional fluid in the absence of dissipative
processes. In \S \ref{sec:QuantumGen}, we discuss the quantum
generation of the scalar and tensor perturbations, and derive
the corresponding power-spectra. The general results are applied
in \S \ref{sec:PBB} to the particular case of the pre-Big Bang
scenario of string cosmology. {}Finally, we discuss our main
results and future applications in \S \ref{sec:Conclusion}.

\section{The evolution equations}
\label{sec:Action}

To keep things as general as possible, we
consider the following  $D$-dimensional action\footnote{
       We adopt the convention $(-,+,\dots,+)$,
       $R_{\nu\rho} = R^{\lambda}_{\;\nu\lambda\rho}$,
       $R^{\mu}_{\;\nu\lambda\rho}=\partial_{\rho}
       \Gamma^{\mu}_{\nu\lambda}+\dots$, and set our units such that
       $\hbar = c = 8 \pi G = 1$.}
\bea
   & & S = \int d^D x \sqrt{-g}
       \Bigl[ {1 \over 2} f(\phi,R)
       - {1 \over 2} \omega (\phi) \phi^{;\mu} \phi_{;\mu}
   \nonumber \\
   & & \hspace{2,65cm}
       - V (\phi) + L^{(c)}  + L_{m} \Bigr],
   \label{e:general-action}
\eea
where $f(\phi,R)$ is an algebraic function of a dimensionless
scalar field $\phi$ and the scalar curvature $R$. $\omega(\phi)$
and $V(\phi)$ are general algebraic functions of $\phi$. Through
the Lagrangian $L^{(c)}$, we allow for the inclusion of terms with
even higher numbers of derivatives such as contracted quadratic
products of the curvature tensor, whereas $L_{m}$ is the
Lagrangian of additional matter fields (e.g. fluids, kinetic
components, axions, moduli, etc.) with its associated
energy-momentum tensor $T_{\mu\nu}$ defined as ${1\over 2}
\sqrt{-g} T^{\mu\nu} \delta g_{\mu\nu} \equiv \delta (\sqrt{-g}
L_{m} )$. By choosing the appropriate parameters, the general
action eq.\ (\ref{e:general-action}) includes Brans-Dicke theory,
non-minimally coupled scalar field, induced gravity,
$R^2$-gravity, {\it etc} \cite{Hwang:1991aj-2001}.
{}For instance, Einstein gravity with a
minimally coupled scalar field corresponds to the case $f=R$,
$\omega = 1$, and $L^{(c)} = 0$.

In a string theory context (see \cite{Lidsey:1999mc} for a recent
review and references therein), the low-energy effective action is
obtained with $f = e^{-\phi} R$, $\omega = - e^{-\phi}$, $V = 0$
and $L^{(c)} = 0$. Higher-order corrections take the form of an
infinite series expansion with expansion parameter
$\alpha'=\lambda_s^2$, where $\lambda_s$ is the fundamental string
length scale. {}For the sake of simplicity, we shall restrict
ourselves to the simplest extension of the lowest-order
gravitational action ensuring that the equations of motion remain
second-order in the fields. In such a case, the most general
Lagrangian density at the next to leading order in the Regge slope
reads \cite{Metsaev:1987zx}:
\bea
   & & L^{(c)} = - {1 \over 2}\alpha^\prime \lambda
       \xi(\phi) \left[ c_1 R_{GB}^2
       + c_2 G^{\mu\nu} \phi_{;\mu} \phi_{;\nu}\right.
   \nonumber \\
   & & \hspace{2,9cm}
      + c_3 \Box \phi \phi^{;\mu} \phi_{;\mu}
      + c_4 (\phi^{;\mu} \phi_{;\mu})^2 \Bigr],
   \label{e:action_corr}
\eea
where $G^{\mu\nu} \equiv R^{\mu\nu}-\frac{1}{2}g^{\mu\nu}R$
is the Einstein tensor and $R_{GB}^2 \equiv R^{\mu\nu\rho\sigma}
R_{\mu\nu\rho\sigma}-4 R^{\mu\nu} R_{\mu\nu} + R^2$ is the
well-know Gauss-Bonnet combination ensuring the ghost-free
character of the theory. In fixing the coefficients $c_i$'s, we
require that the full action agrees with the three-graviton
scattering amplitude \cite{Metsaev:1987zx}, and thus impose the
constraints, $\xi = -e^{-\phi}$,  $c_3 = -[c_2 + 2 (c_1 + c_4)]/2$
and $c_1 = -1$, working in units $\alpha'=1$. $\lambda$ is an
additional parameter allowing for different species of string
theories, $\lambda = -1/4,-1/8$ for the Bosonic and Heterotic
string respectively, and $\lambda =0$ for superstrings. The
inclusion of such higher-order corrections for the background
evolution has recently been investigated in the context of the
pre-Big Bang scenario of string cosmology, leading to a number of
promising non-singular cosmologies which smoothly interpolate
between the growing to the decreasing curvature regime
\cite{pbb-higher-corr,Cartier:1999vk}.

By varying eq.\ (\ref{e:general-action}) with respect to the
metric and the scalar field we can derive the gravitational field
equation and the equation of motion for the scalar field $\phi$,
respectively:
\bea
   & & {F} G^{\mu}_{\nu} =
       \omega \left( \phi^{;\mu} \phi_{;\nu}
       - {1 \over 2} \delta^{\mu}_{\nu}
       \phi^{;\rho} \phi_{;\rho} \right)
       - {1 \over 2} \delta^{\mu}_{\nu}
       \left( {F} R - f + 2 V \right)
   \nonumber \\
   & & \hspace{1,2cm}
       + {F}^{;\mu}_{\;\;\;\;\nu}
       - \delta^{\mu}_{\nu} \Box {F}
       + T^{(c)\mu}_{\;\;\;\;\;\nu} + T^{\mu}_{\nu},
   \label{e:GFE} \\
   & & \Box \phi + {1 \over 2\omega} \left( f_{,\phi}
       + \omega_{,\phi}\phi^{;\mu} \phi_{;\mu} - 2 V_{,\phi}
       - T^{(c)}_{\phi} \right) = 0,
   \label{e:EOM}
\eea
where $F \equiv {\partial f \over \partial R}$.
$T^{(c)\mu}_{\;\;\;\;\;\nu}$ and $T^{(c)}_{\phi}$ represent the
contributions derived from the next to leading order corrections
given by eq.\ (\ref{e:action_corr}),
\bea
   & & T^{(c)\mu}_{\;\;\;\;\;\nu} \equiv - \alpha^\prime \lambda \Bigg\{
       - 4 c_1 \Bigl[ \left(R^{\mu}_{\;\;\sigma\nu\tau} +R_{\nu\sigma}
       \delta^{\mu}_{\tau}-\delta^{\mu}_{\nu}R_{\sigma\tau}
       \right)\xi^{;\sigma\tau}
   \nonumber \\
   & & \quad\quad\quad
       +G^{\mu\sigma} \xi_{;\nu\sigma} - G^{\mu}_{\nu}\Box \xi \Bigr]
       + c_1 \xi\, \aleph^\mu_\nu
   \nonumber \\
   & & \quad
       + c_2 \Bigl\{\xi \Bigl[(\delta^{\mu}_{\nu}R_{\sigma\tau}
       -R^{\mu}_{\;\;\sigma\nu\tau})\phi^{;\sigma}\phi^{;\tau}
       -R^\mu_\sigma \phi_{;\nu} \phi^{;\sigma}
   \nonumber \\
   & & \quad\quad\quad
       -R^\sigma_\nu \phi^{;\mu} \phi_{;\sigma}\Bigr]
   \nonumber \\
   & & \quad\quad\quad
       +{1\over2} \xi \left[G^{\mu}_{\nu}\phi^{;\sigma} \phi_{;\sigma}
       + R \phi^{;\mu}\phi_{;\nu}-2
       \phi^{;\mu\sigma}\phi_{;\nu\sigma}\right]
   \nonumber \\
   & & \quad\quad\quad
       +{1\over2}\xi\left[ 2\Box
       \phi\;\phi^{;\mu}_{\;\;\;\nu}-\delta^{\mu}_{\nu}(\Box \phi)^2
       +\delta^{\mu}_{\nu}\phi^{;\sigma\tau}\phi_{;\sigma\tau} \right]
   \nonumber \\
   & & \quad\quad\quad
       +{1\over2}\left[\delta^{\mu}_{\nu}\left(\Box \xi
       \,\phi^{;\sigma}\phi_{;\sigma}
       -\xi^{;\sigma\tau}\phi_{;\sigma}\phi_{;\tau}\right)
       -\Box \xi \,\phi^{;\mu}\phi_{;\nu}\right]
   \nonumber \\
   & & \quad\quad\quad
       +{1\over2}\left[\xi^{;\mu\sigma}\phi_{;\nu}\phi_{;\sigma}
       +\xi_{;\nu\sigma}\phi^{;\mu}\phi^{;\sigma}
       -\xi^{;\mu}_{\;\;\;\nu}\phi^{;\sigma}\phi_{;\sigma}\right]
   \nonumber \\
   & & \quad\quad\quad
       +{1\over2}\left[\Box\phi(\xi^{;\mu}\phi_{;\nu}
       +\xi_{;\nu}\phi^{;\mu})
       +2\xi^{;\sigma}\phi_{;\sigma}\phi^{;\mu}_{\;\;\;\nu} \right]
   \nonumber \\
   & & \quad\quad\quad
       +{1\over2}\left[2\delta^{\mu}_{\nu}\left(\xi^{;\sigma}
       \phi^{;\tau}\phi_{\;\sigma\tau}
       -\Box\phi\,\xi^{;\sigma}\phi_{;\sigma}\right)
       -\xi_{;\sigma}\phi^{;\mu\sigma}\phi_{;\nu}\right]
   \nonumber \\
   & & \quad\quad \quad
       -{1\over2}\left[\xi^{;\sigma}\phi^{;\mu}\phi_{;\nu\sigma}
       +\xi^{;\mu}\phi_{;\nu\sigma}\phi^{;\sigma}
       +\xi_{;\nu}\phi^{;\mu\sigma}\phi_{;\sigma}\right] \Bigr\}
   \nonumber \\
   & & \quad
       + c_{3} \Bigl\{\xi \left[\phi^{;\mu\rho} \phi_{;\nu}\phi_{;\rho}
       +\phi^{;\mu} \phi_{;\nu}^{\;\;\;\rho}\phi_{;\rho}\right]
   \nonumber \\
   & & \quad\quad\quad
       -\xi \left[\delta^{\mu}_{\nu} \phi^{;\rho\sigma}
       \phi_{;\rho}\phi_{;\sigma}
       \Box \phi \,\phi^{;\mu} \phi_{;\nu} \right]
   \nonumber \\
   & & \quad\quad\quad
       +{1 \over 2} \phi^{;\rho} \phi_{;\rho} \left[\xi^{;\mu} \phi_{;\nu}
       + \xi_{;\nu} \phi^{;\mu} - \delta^{\mu}_{\nu} \xi^{;\rho}
       \phi_{;\rho} \right] \Bigr\}
   \nonumber \\
   & & \quad
       + {1 \over 2}c_{4} \xi \phi^{;\sigma} \phi_{;\sigma} \left[
       \delta^\mu_\nu\phi^{;\rho} \phi_{;\rho}
       -4 \phi^{;\mu} \phi_{;\nu} \right]\Bigg\},
   \label{e:GFE_corr} \\
   & & \aleph^\mu_\nu \equiv {1\over2}\delta^\mu_\nu R^{2}_{GB}
        + 4 R^\mu_{\;\;\sigma\nu\tau} R^{\sigma\tau} - 2 R R^\mu_\nu
    \nonumber \\
   & & \quad \quad\quad
       - 2 R^{\mu}_{\;\;\rho\sigma\tau}R_\nu^{\;\;\rho\sigma\tau}
       + 4 R^\mu_\sigma R_\nu^\sigma,
   \label{e:aleph} \\
   & & T^{(c)}_{\phi} \equiv \alpha^\prime \lambda \Biggl\{
       c_1 \xi_{,\phi} R^2_{GB}
   \nonumber \\
   & & \quad + c_2 G^{\mu\nu} \left( \xi_{,\phi} \phi_{;\mu} \phi_{;\nu}
       - 2 \xi_{;\mu} \phi_{;\nu} -2 \xi \phi_{;\mu\nu} \right)
   \nonumber \\
   & & \quad
       + c_3 \Bigl\{ \left( \xi_{,\phi} \Box\phi + \Box \xi \right)
       \phi^{;\mu} \phi_{;\mu}
       - 2 \Box \phi \xi^{;\mu} \phi_{;\mu}
       + 4 \xi^{;\mu} \phi^{;\nu} \phi_{;\mu\nu}
   \nonumber \\
   & & \quad\quad\quad
       +2 \xi \Bigl[ \phi^{;\mu\nu} \phi_{;\mu\nu}
       + R^{\mu\nu}\phi_{;\mu}\phi_{;\nu}
       -(\Box \phi)^2 \Bigr] \Bigr\}
   \nonumber \\
   & & \quad
       + c_4 \Bigl[ \xi_{,\phi} (\phi^{;\mu} \phi_{;\mu})^2
       -4 \xi^{;\mu} \phi_{;\mu} \phi^{;\nu} \phi_{;\nu} - 4 \xi
       \,\Box \phi \,\phi^{;\mu} \phi_{;\mu}
   \nonumber \\
   & & \quad\quad\quad
       -8 \xi \phi^{;\mu} \phi^{;\nu}\phi_{;\mu\nu}
       \Bigr] \Biggr\}.
   \label{e:EOM_corr}
\eea
Equations (\ref{e:GFE})--(\ref{e:EOM_corr}) form a complete
set of covariant generalised Einstein equations which include
higher-order corrections. Hence they extend the domain of validity
of any solutions into the highly-curved regimes included in eq.\
(\ref{e:action_corr}). This system of equations provides the
framework to study both the cosmological background and evolution
of metric and field perturbations. Note that $\aleph^\mu_\nu$
corresponds to a quadratic expression which vanishes in four
dimensions on account of the algebraic identities satisfied by the
Riemann tensor \cite{Field_DeWitt}.

\section{Perturbed field equations}
\label{sec:Perturbation}

{}From now on, we concentrate on the four-dimensional case, and
consider the metric of a spatially homogeneous and isotropic model
with the most general perturbations:
\bea
   & & d s^2 = - a^2 \left( 1 + 2 \alpha\right) d \eta^2
       - a^2 \left( \beta_{,i} + B_i \right) d \eta d x^i
    \label{e:perturbed-metric} \\
   & & \quad
       + a^2 \Big[ g^{(3)}_{ij} \left(1+2\varphi\right)
       + 2 \gamma_{,i|j} + 2 C_{(i|j)}
       + 2 C_{ij} \Big] d x^i d x^j,
    \nonumber
\eea
where $a(t)$ is the cosmic scale factor with $dt \equiv a d
\eta$. Latin letters denote space indices. $\alpha ({\bf x}, t)$,
$\beta ({\bf x}, t)$, $\varphi ({\bf x}, t)$ and $\gamma ({\bf x},
t)$ characterise the scalar-type perturbation, $B_i ({\bf x}, t)$
and $C_i ({\bf x}, t)$ are transverse ($B^i_{\;\;|i} = 0 =
C^i_{\;\;|i}$) and represent the vector-type perturbation, whereas
$C_{ij} ({\bf x}, t)$ is transverse and tracefree ($C^j_{i|j} = 0
= C^i_i$), and corresponds to the tensor-type perturbation.
Indices are based on $g^{(3)}_{ij}$ as the metric, and a vertical
bar indicates a covariant derivative based on $g^{(3)}_{ij}$.

We decompose the energy-momentum tensor of the additional matter
and the dilaton field into
\bea
&& T^\mu_\nu  ({\bf x}, t) = \bar T^\mu_\nu (t)
   + \delta T^\mu_\nu ({\bf x}, t), \\
&& \phi ({\bf x}, t) = \bar \phi (t) + \delta \phi ({\bf x}, t).
\eea
An overbar indicates a background order quantity and will be
omitted unless necessary. The three types of perturbations
decouple from each other due to the symmetry in the background
field equations and the fact that we are working to linear order
in the perturbations. Thus, we can handle them individually and we
find it convenient to separate their respective contributions in
the perturbed energy-momentum tensor. We use the superscripts
$(s)$, $(v)$ and $(t)$ for the scalar, vector and tensor parts,
respectively,  such that $ \delta T^\mu_\nu \equiv \delta
T^{(s)\,\mu}_{\;\;\;\;\;\,\nu} + \delta
T^{(v)\,\mu}_{\;\;\;\;\;\,\nu} + \delta
T^{(t)\,\mu}_{\;\;\;\;\;\nu}$. Using a normalised $(u^\mu
u_\mu=-1)$ four-vector and its associated projection tensor
$h_{\mu\nu} \equiv g_{\mu\nu}+u_\mu u_\nu$, a covariant
decomposition of the (imperfect fluid) energy-momentum tensor into
fluid quantities is given by \cite{covariant-T}
\bea
  T_{\mu\nu} = \rho u_\mu u_\nu + p h_{\mu\nu}  +
  q_\mu u_\nu + q_\nu u_\mu +\pi_{\mu\nu},
\label{e:Energy-momentum}
\eea
where the background is assumed to be made up of a perfect fluid.
Here, we have defined $\rho \equiv T_{\mu\nu} u^\mu u^\nu$ as the
energy density and $p \equiv \frac{1}{3} T_{\mu\nu} h^{\mu\nu}$ as
the pressure. The energy flux and anisotropic pressure are,
respectively, $q_\mu \equiv - T_{\nu\sigma} u^\nu h^\sigma_\mu$
and $\pi_{\mu\nu}\equiv T_{\sigma\tau} h^\sigma_\mu h^\tau_\nu$,
and satisfy $q_\mu u^\mu = 0 = \pi_{\mu\nu} u^\mu$,
$\pi_{\mu\nu}=\pi_{\nu\mu}$. Convenient decompositions for the
fluid four-velocity and the energy flux are, respectively, $u_\mu
\equiv -a(1+\alpha,\delta u_i)$ and $q_\mu \equiv a (0,\delta
q_i)$. The content of the energy-momentum tensor to linear order
then becomes 
\bea 
&&  T^0_0 = -\left(\rho +\delta\rho\right),
    \label{e:EM_00} \\
&&  T^0_i = -\left(\rho + p\right) \delta u_i + \delta q_i,
    \label{e:EM_0i} \\
&&  T^i_j = \left(p+\delta p\right)\delta^i_j
    + \pi^i_j.
    \label{e:EM_ij}
\eea
In general, the decomposition of the flux in eq.\
(\ref{e:Energy-momentum}) suggests the apparition of additional
degrees of freedom. For instance, we may choose to consider the
normal frame $\delta u_i = 0$ or the energy frame $\delta q_i =
0$. In linear perturbation theory, however, $\delta u_i$ and
$\delta q_i$ always appear together in a frame independent
combination of the from of eq.\ (\ref{e:EM_0i}). Hence, we can
investigate the gauge transformation properties of such a
combination in a frame independent way \cite{Hwang:1991aj-2001}.
{}For later convenience, we
split eqs.\ (\ref{e:EM_0i},\ref{e:EM_ij}) into the three-types
of perturbations as
\bea
&&  T^0_i \equiv \left(\rho + p\right)
    \left( v^{(s)}_{i}+ v^{(v)}_{i} \right),
    \label{e:EM_0i_bis} \\
&&  T^i_j \equiv \left(p+\delta p\right)\delta^i_j
    + \pi^{(s)i}_{\;\;\;\;\,j} + \pi^{(v)i}_{\;\;\;\;\,j}
    + \pi^{(t)i}_{\;\;\;\;j}.
    \label{e:EM_ij_bis}
\eea

The background equations in \S \ref{ssec:BG} are presented
considering the general spatial curvature $K$ of the background,
whereas for the perturbation equations in \S
\ref{ssec:Scalar}-\ref{ssec:Tensor} we assume a flat background
(i.e. $K = 0$). For the scalar-type perturbation we set
$T^{(s)\,\mu}_{\;\;\;\;\;\,\nu} = 0$, whereas for the vector and
tensor type perturbations we keep the contribution from the fluid
energy-momentum tensor.

\subsection{Background field equations}
\label{ssec:BG}

{}From eqs. (\ref{e:GFE})--(\ref{e:EOM}) we obtain a set of
equations describing the background evolution of the
homogeneous-isotropic FLRW world model in our generalised gravity
model:
\bea
   & & H^2 + {K \over a^2}
       = {1 \over {6 F}} \Big( \omega \dot{\phi^2}
       + R {F} - f + 2 V-6 H \dot{F}
   \nonumber \\
   & & \hspace{2,6cm}
       - 2 T^0_0 - 2 T^{(c)0}_{\;\;\;\;\,0} \Big),
   \label{e:BG1} \\
   & & \dot H - {K \over a^2}
       = {1 \over 2{F}} \Big( - \omega \dot{\phi^2}
       + H \dot{{F}} -\ddot{{F}} + T^0_0
       - {1 \over 3} T^i_i
   \nonumber \\
   & & \hspace{2,6cm}
       + T^{(c)0}_{\;\;\;\;\,0}
       - {1 \over 3} T^{(c)i}_{\;\;\;\;\,i} \Big),
   \label{e:BG2} \\
   & & \ddot{\phi} + 3 H \dot{\phi} + \frac{1}{2\;\omega}
       \left( \omega_{,\phi} \dot{\phi}^2 - f_{,\phi}
       + 2 V_{,\phi} + T^{(c)}_{\phi} \right) = 0,
   \label{e:BG-EOM}
\eea
where we denote the Hubble parameter as $H \equiv \dot a/a$,
the Ricci scalar $R = 6 (2 H^2 + \dot H + K/a^2)$ and the
higher-order contributions are given by
\bea
   & & T^{(c)0}_{\;\;\;\;\; 0} = -\alpha^\prime \lambda \left[
       12 c_1 \dot{\xi} H \left( H^2 + {K \over a^2} \right)
       \right.
   \nonumber \\
   & & \quad
       - {3\over2} c_2 \xi \dot{\phi}^2
       \left( 3H^2+ {K \over a^2} \right)
   \nonumber \\
   & & \quad
       \left. + {1\over2} c_3 \dot{\phi}^3
       \left( \dot{\xi} -6 \xi H \right)
       - {3\over2} c_4 \xi \dot{\phi}^4 \right],
   \\
   & & T^{(c)i}_{\;\;\;\;\,i} =
       -3 \alpha^\prime \lambda \Biggl\{
       4c_1 \left[ \ddot{\xi} \left( H^2 + {K \over a^2} \right)
       + 2 \dot{\xi} H \left( \dot{H} +H^2 \right) \right]
   \nonumber \\
   & & \quad
       - \frac{1}{2} c_{2}\dot{\phi}\left[\xi \dot{\phi}
       \left(2 \dot{H}+ 3 H^2 - {K \over a^2} \right)
       + 4 \xi \ddot{\phi} H + 2\dot{\xi} \dot{\phi}H \right]
   \nonumber \\
   & & \quad
       - {1\over2} c_3 \dot{\phi}^2 \left( 2\xi\ddot{\phi}
       + \dot{\xi}\dot{\phi} \right)
       + {1 \over 2} c_4 \xi \dot{\phi}^4 \Biggr\},
   \\
   & & T^{(c)}_{\phi}
       = \alpha^\prime \lambda \Biggl\{
       24 c_1 \xi_{,\phi} \left( \dot{H} + H^2 \right)
       \left( H^2 + {K \over a^2} \right)
   \nonumber \\
   & & \quad
       + 3 c_2 \left[ - \left( H^2+ {K \over a^2} \right)
       \left( \dot{\xi} \dot{\phi} +2\xi\ddot{\phi} \right)
       \right.
   \nonumber \\
   & & \hspace{1,3cm}
       \left. -2 \xi \dot{\phi} H \left(2\dot{H} +3H^2
       + {K \over a^2} \right) \right]
   \nonumber \\
   & & \quad
       + c_3 \dot{\phi} \Bigl[ \dot{\phi} \ddot{\xi}
       +3\dot{\xi}\ddot{\phi}
       -6 \xi \left( \dot{\phi} \dot{H} + 2 \ddot{\phi} H
       + 3 \dot{\phi} H^2 \right) \Bigr]
   \nonumber \\
   & & \quad
       + c_4 \dot{\phi}^2 \left(
       - 3 \dot{\xi} \dot{\phi}-12\xi\ddot{\phi}
       - 12 \xi \dot{\phi} H \right) \Biggr\}.
   \label{e:BG-EOM_corr}
\eea
The solutions of the system of equations
(\ref{e:BG1})--(\ref{e:BG-EOM_corr}) provide the cosmological
``graviton-scalar field'' background in which we will study the
propagation of scalar field and metric perturbations. Equations
(\ref{e:BG1})--(\ref{e:BG-EOM}) are not independent. We can derive
eq.\ (\ref{e:BG2}) from eqs.\ (\ref{e:BG1},\ref{e:BG-EOM}), by
using $\dot T^0_0 + H ( 3 T^0_0 - T^i_i ) = 0$ which follows from
energy-momentum conservation of the additional fluids or fields.

\subsection{Scalar-type perturbations}
\label{ssec:Scalar}

To investigate the scalar-type perturbations there are many
different temporal gauge conditions available for us to use. Due
to the homogeneity of the background three-space the spatial gauge
transformation is trivial \cite{Pert_Bardeen}: although $\beta$
and $\gamma$ change under the spatial gauge transformation, they
always appear together in a spatially gauge-invariant combination
$\chi \equiv a (\beta + a \dot \gamma)$, whereas $\alpha$ and
$\varphi$ are spatially gauge-invariant. Except for the
synchronous gauge condition, which happens to be the most widely
used one, all the other fundamental temporal gauge conditions fix
the gauge degrees of freedom completely, thus a variable in such a
gauge condition uniquely corresponds to a gauge-invariant
combination of the variable concerned and the variable used in
fixing the gauge condition, see eq.\ (\ref{GI-variable}) and
below. The most suitable gauge condition depends on the type of
the problem and in general is not known {\it a priori}, hence the
gauge-ready method proposed in
\cite{Pert_Bardeen,Hwang:1991aj-2001} appears particularly
convenient since it allows for a flexible use of the various
fundamental gauge conditions.

In handling the perturbations involved with the scalar field in
Einstein or generalised gravity theories [without the $L^{(c)}$
term in eq. (\ref{e:action_corr})], it is known that the field
fluctuation in the uniform-curvature gauge or equivalently the
curvature fluctuation in the uniform-field gauge (see below) allow
the simplest analysis \cite{Hwang-MSF}. Following
\cite{Hwang:1999gf}, we derive an equation for a gauge-invariant
combination $\varphi_{\delta \phi}$ defined as
\bea
   & & \varphi_{\delta \phi} \equiv \varphi
       - {H \over \dot \phi} \delta \phi
       \equiv - {H \over \dot \phi} \delta \phi_\varphi.
   \label{GI-variable}
\eea
In the uniform-field gauge, which takes $\delta \phi \equiv 0$,
the gauge-invariant quantity $\varphi_{\delta \phi}$ is identified
with $\varphi$. Similarly $\delta \phi_\varphi$ is the same as
$\delta \phi$ in the uniform-curvature gauge which sets $\varphi = 0$.
The gauge invariant
combination $\varphi_{\delta \phi}$ was first introduced by Lukash
in \cite{Lukash-1980,Mukhanov-1988}. If we have a solution in one
gauge condition, the other solutions in the same gauge as well as
the ones in other gauges can be derived from it as linear
combinations. In the present case, choosing the uniform-field
gauge also implies $\delta \xi = 0$. We consider a case with $F =
F (\phi)$, thus we have $\delta F = 0$. Ignoring the additional
matter, that is we do not include any other components other than
the scalar (dilaton) field, i.e. $T^{(s)\mu}_{\;\;\;\;\;\nu} = 0$,
and following the same steps described above eq.\ (3) in
\cite{Hwang:1999gf}, we can derive a closed form equation
describing the classical evolution of the scalar-metric
perturbation. Explicitly, we find
\bea
   & & \frac{1}{a^3 Q^{(s)}} \left(a^3
       Q^{(s)} \dot\varphi_{\delta \phi} \right)^\cdot -
       s^{(s)} \frac{\Delta}{a^2}\varphi_{\delta \phi} = 0,
   \label{varphi-eq}
\eea
where
\bea
   & & Q^{(s)} \equiv \frac{\omega\dot\phi^2
       +3\frac{\left(\dot{\cal F}+Q^{(s)}_{a}\right)^2}{2{\cal
       F}+Q^{(s)}_{b}}+Q^{(s)}_{c}}
       {\left(H+\frac{\dot{\cal F}+Q^{(s)}_{a}}{2{\cal F}
       +Q^{(s)}_{b}}\right)^{2}},
   \nonumber \\
   & & s^{(s)} \equiv 1 + \frac{Q^{(s)}_{d}
       +\frac{\dot{\cal F}+Q^{(s)}_{a}}{2{\cal F}
       +Q^{(s)}_{b}}Q^{(s)}_{e}+\left(\frac{\dot{\cal F}
       +Q^{(s)}_{a}}{2{\cal F}+Q^{(s)}_{b}}\right)^2 Q^{(s)}_{f}}
       {\omega\dot\phi^2+3\frac{\left(\dot{\cal F}
       +Q^{(s)}_{a}\right)^2}{2{\cal F}+Q^{(s)}_{b}}+Q^{(s)}_{c}}.
   \label{e:s_scalar}
\eea
The contribution arising from the higher-order corrections is
summarised in the quantities
\bea
   & & Q^{(s)}_{a} \equiv \alpha^\prime \lambda
       \left[ - 4 c_1 \dot \xi H^2 + 2 c_2 \xi \dot \phi^2 H
       + c_3 \xi \dot \phi^3 \right],
   \nonumber \\
   & & Q^{(s)}_{b} \equiv \alpha^\prime \lambda
       \left[ - 8 c_1 \dot \xi H + c_2 \xi \dot \phi^2 \right],
   \nonumber \\
   & & Q^{(s)}_{c} \equiv \alpha^\prime \lambda \dot \phi^2
       \left[ -3 c_2 \xi H^2 + 2 c_3 \dot \phi \left( \dot \xi
       - 3 \xi H \right) - 6 c_4 \xi \dot \phi^2 \right],
   \nonumber \\
   & & Q^{(s)}_{d} \equiv \alpha^\prime \lambda \dot \phi^2
       \Big[ - 2 c_2 \xi \dot H - 2 c_3 \left( \dot \xi \dot \phi
       + \xi \ddot \phi - \xi \dot \phi H \right)
   \nonumber \\
   & & \hspace{2,1cm}
       + 4 c_4 \xi \dot \phi^2 \Big],
   \nonumber \\
   & & Q^{(s)}_{e} \equiv \alpha^\prime \lambda
       \Big[ - 16 c_1 \dot \xi \dot H + 2 c_2 \dot \phi
       \left( \dot \xi \dot \phi + 2 \xi \ddot \phi
       - 2 \xi \dot \phi H \right)
   \nonumber \\
   & & \hspace{2,1cm}
       - 4 c_3 \xi \dot \phi^3 \Big],
   \nonumber \\
   & & Q^{(s)}_{f} \equiv \alpha^\prime \lambda
       \left[ 8 c_1 \left( \ddot \xi - \dot \xi H \right)
       + 2 c_2 \xi \dot \phi^2 \right].
   \label{e:Qf_scalar}
\eea
In units $\alpha^\prime=1$, our result reproduces those previously
obtained for the particular case $c_1 = -1$, $c_2 = c_3 = c_4 = 0$
and $\lambda = -1/4$\cite{Hwang:1999gf}.

\subsection{Vector-type perturbations}
\label{ssec:Vector}

Rotational perturbations do not enter the $(0,0)$ components of
the field equations, but they do contribute to the $(0,i)$ and
$(i,j)$ components. In fact, our scalar field perturbation does
not couple directly with these vector-type perturbations. To
investigate the influence of the scalar field, we consider an
additional fluid present in the system. We can introduce the
vector-type fluid quantities of the additional matter as
\bea
   & & \delta T^{(v)0}_{\;\;\;\;\;i}
       \equiv ( \rho + p) v^{(v)} Y^{(v)}_i, \quad
       \delta T^{(v)i}_{\;\;\;\;\;j}
       \equiv \pi^{(v)} Y^{(v)i}_{\;\;\;\;\;j},
   \label{e:vec_ij}
\eea
where $Y^{(v)}_i$ and $Y^{(v)i}_{\;\;\;\;\;j}$ are vector-type
harmonic functions \cite{Bardeen:1980kt}. $v^{(v)}$ is the
velocity variable related to the vorticity and $\pi^{(v)}$ is the
anisotropic stress. In the field equations, the vector
perturbations always appear in a gauge-invariant combination $B_i
+ a \dot C_i \equiv \Psi^{(v)} Y^{(v)}_i$. From the $(0,i)$ and
$(i,j)$ components of the gravitational field equation we can
derive\footnote{
       Compared with \cite{Bardeen:1980kt} we have
       $v^{(v)} \equiv v_c$, $\pi^{(v)} \equiv p \pi^{(1)}_T$
       and $\Psi^{(v)} = \Psi$.}:
\bea
   & & \frac{k^2}{2 a^2} Q^{(v)} \Psi^{(v)}
       = \left( \rho + p \right) v^{(v)},
   \label{rot-eq1} \\
   & & \frac{1}{a^4} \big[ a^4 \left( \rho + p \right) v^{(v)} \big]^\cdot
       = - {k \over 2 a} \pi^{(v)},
   \label{rot-eq2}
\eea
where we have introduced
\bea
   Q^{(v)} \equiv F - \frac{1}{2} \alpha^\prime \lambda
   \left( 8 c_1 \dot \xi H - c_2 \xi \dot \phi^2 \right).
   \label{e:vector_Q}
\eea
If we ignore the anisotropic stress of the fluid, $\pi^{(v)}$,
eq.\ (\ref{rot-eq2}) implies the conservation of the
angular-momentum of the fluid, as
\bea
   a^3 \left( \rho + p \right) \cdot a
   \cdot v^{(v)} ({\bf x}, t) = L({\bf x}).
   \label{rot-sol}
\eea
Notice that this result is independent of the generalised nature
of our gravity model. The evolution of vorticity depends on our
generalised gravity indirectly through the background evolution.
The generalised nature, however, appears directly in connecting the
vorticity to the metric perturbation in eq. (\ref{rot-eq1}).
Equation (\ref{rot-eq2}), in fact, follows
from the $0$-component of the conservation of the fluid part of
the energy-momentum tensor ($T^{0;\nu}_\nu = 0$), hence is
naturally independent of the field equation. Conservation of
angular momentum forces the rotational perturbation to decay in an
expanding medium, which renders it generally cosmologically
uninteresting in the context of linear perturbations in expanding
universe.

\subsection{Tensor-type perturbations}
\label{ssec:Tensor}

{}From the $(i,j)$ component of the field equation we can derive the
linearised equation for the gravitational wave perturbation
\bea
   & & {1 \over a^3 Q^{(t)}} \left( a^3 Q^{(t)} \dot{C}^{i}_{j}
       \right)^\cdot - s^{(t)} {\Delta \over a^2} C^{i}_{j}
       = {1 \over Q^{(t)}} \delta T^{(t)i}_{\;\;\;\;\,j},
   \label{GW-eq}
\eea
where
\bea
   & & Q^{(t)} \equiv {F} - {1 \over 2} \alpha^\prime \lambda
       \left( 8 c_1 \dot{\xi} H -c_2\xi \dot{\phi}^2 \right),
   \nonumber \\
   & & s^{(t)} \equiv {1 \over Q^{(t)}}
       \left[ {F} - {1 \over 2} \alpha^\prime \lambda
       \left( 8 c_1 \ddot{\xi} + c_2\xi \dot{\phi}^2 \right) \right].
       \label{e:s_tensor}
\eea
$\delta T^{(t)i}_{\;\;\;\;\,j}$ includes contributions to the
tensor-type energy-momentum tensor, i.e., transverse-tracefree
anisotropic stresses which can possibly arise if we include
additional imperfect fluids. Since the gauge transformation does
not affect the transverse-tracefree parts, $C^i_j$ and $\delta
T^{(t)i}_{\;\;\;\;\,j}$ are naturally gauge invariant. In the
context of string cosmology, $F=-\omega=-\xi=e^{-\phi}$ and $V=0$,
we note that one easily recovers the $\alpha'$ corrections
obtained in \cite{Cartier:2001gc}.

\section{Quantum generation and evolution of perturbations}
\label{sec:QuantumGen}

\subsection{Classical evolution}
\label{ssec:classical evolution}

Assuming that there is no additional matter, $T^\mu_\nu = 0$, we
have been able to derive a closed form equation for the linearised
classical evolution of both scalar and tensor metric
perturbations. In a unified formalism \cite{Hwang-Noh-2001-fR},
the dynamics of a perturbed
variable $\Phi$ is typically governed by the wave equation,
\bea
   \frac{1}{a^3 Q} \left( a^3 Q \dot \Phi \right)^\cdot
   - s \frac{\Delta}{a^2} \Phi= 0,
   \label{e:unified-wave-cs}
\eea
which may be readily derived from the second-order perturbed action
\bea
   \delta^{(2)} S = \frac{1}{2} \int  a^3 Q \left(\dot \Phi^2
   - \frac{s}{a^2} \Phi^{|i}\Phi_{,i}\right) d^3 x dt,
   \label{e:action-2order}
\eea
where $\left(Q,s,\Phi\right) \in \left\{
\left(Q^{(s)},s^{(s)}, \varphi_{\delta\phi}\right);
\left(Q^{(t)},s^{(t)}, C^i_j\right)\right\}$. Introducing the
quantities $\psi \equiv z \Phi$ with $z\equiv a\sqrt{Q}$, the
linearised wave equation (\ref{e:unified-wave-cs}) can be written
for each Fourier mode, $\psi_k \equiv z \Phi_k$, in terms of the
eigenstates of the Laplace-Beltrami operator, $\Delta \psi_k = -
k^2 \psi_k$. Explicitly, eq.\ (\ref{e:unified-wave-cs}) becomes
\bea
   \psi^{\prime\prime}_k + \Bigl[s k^2 - V(\eta) \Bigr] \psi_k = 0,
   \quad
   V(\eta) = \frac{z^{\prime\prime}}{z},
   \label{e:unified-wave-cf}
\eea
where $~^\prime \equiv d/d\eta$. The ``pump'' field $z$ accounts for
the parametric amplification of the metric fluctuations, whereas
$s$ may be interpreted as an effective shift in the frequency.
We note that the wave equation (\ref{e:unified-wave-cf}) can be
deduced from the second-order perturbed action
\bea
   \delta^{(2)}{\cal S} = \frac{1}{2} \int
   \left[ {\psi^\prime}^{2} + \frac{z^{\prime\prime}}{z} \psi^2
   + s \psi \Delta \psi \right] d^{3}x d\eta,
   \label{e:unified-action-cf}
\eea
where the kinetic term is diagonal, emphasising the canonical
nature of the variable $\psi$. To linear order, the decomposition
in Fourier modes implies that each comoving wavenumber $k$ evolves
independently of other comoving modes, satisfying eq.\
(\ref{e:unified-wave-cf}) for all times. In a General Relativity
context and in spatially flat FLRW manifold, we recall that eq.\
(\ref{e:unified-wave-cf}) reduces to the equation of a minimally
coupled massless scalar field for tensor perturbations
\cite{Grishchuk:1975ny}, whereas a time-dependent effective mass
is present in the scalar perturbation case \cite{Hwang:1993cv}. In
the present context of generalised gravity theory, however, the
wave equation differs from the Klein-Gordon equation for two
reasons. On the one hand, both scalar and tensor perturbations are
coupled not only to the background scale factor, but potentially
also to the scalar field background through the algebraic function
$F = \partial f(\phi,R)/\partial R$. Hence, the growth of the
comoving amplitude of metric perturbations rises because of the
joint contribution of the metric and the scalar field background
to the pump field $z=z(a,\phi)$. On the other hand, the
higher-order corrections act as an additional source for the
parametric amplification of the metric perturbations. These
deviations from the General Relativity case manifest themselves
through the unique function $Q$, which encompasses all sources of
modification.

In the case where the background evolution undergoes
power-law-type inflation, the external potential responsible for
the parametric amplification process
\cite{Grishchuk:1990bj-1992tw} behaves as $V(\eta) \sim
|\eta|^{-2}$, where we have set $z\sim |\eta|^\gamma$. For each
comoving wavenumber, the wave equation (\ref{e:unified-wave-cf})
then reduces to
\bea
   \psi^{\prime\prime}_k + k^2\left[\pm \tilde s
   - \frac{\nu^2-1/4}{|k\eta|^2} \right] \psi_k = 0,
   \label{e:unified-Bessel-wave}
\eea
where we have introduced $\tilde s = |s|$ and $\nu \equiv
\frac{1}{2}\left|1-2 \gamma \right|$. At lowest-order, i.e.
neglecting the frequency shift and corrections in the pump field,
the difference between scalar and tensor metric perturbations is
encoded in the argument $\nu \in \{\nu_s;\nu_t\}$, reflecting the
background dynamics through the exponent $\gamma \in
\{\gamma_s;\gamma_t\}$ of the pump field. The dynamical behaviour
of the canonical variable, given by eq.\
(\ref{e:unified-Bessel-wave}), can be divided into two asymptotic
regimes. On the one hand, the power-law behaviour of the external
potential implies that it vanishes in the asymptotic past ($|k
\eta| \gg 1$). Hence, eq.\ (\ref{e:unified-Bessel-wave}) yields
two oscillating solutions, corresponding to negative and positive
frequency modes respectively. On the other hand, the monotonic
growth of the effective potential of the perturbation ensures that
it will soon drive the evolution of the metric perturbation
(parametric amplification process). In the large wavelength
$(|k\eta| \ll 1)$ limit, which corresponds to scales larger than
the Hubble radius, eq.\ (\ref{e:unified-Bessel-wave}) yields the
solution
\bea
    \Phi_k (\eta,{\bf x}) &=& C_k ({\bf x})
    + D_k ({\bf x}) \int^\eta \frac{d\eta^\prime}{z^2}
    \nonumber \\
&=& C_k ({\bf x}) + D_k ({\bf x}) |\eta|^{1-2\gamma},
    \label{e:unified-wave-sol}
\eea
where $C_k({\bf x})$ and $D_k({\bf x})$ are constants of
integration. For $\gamma < \frac{1}{2}$, the metric perturbation
$\Phi_k$ approaches a constant for $\eta \rightarrow 0_{-}$. For
$\gamma \geq \frac{1}{2}$ however, the second term with $D_k$
grows in time, with an additional logarithmic factor appearing at
$\gamma = \frac{1}{2}$. More generally, for both scalar and tensor
metric perturbations in the large-scale limit, $ s k^2 \ll
z^{\prime\prime}/z$, thus ignoring the Laplacian term in eq.\
(\ref{e:unified-wave-cf}), the exact solution is given by
\bea
    \Phi_k (\eta,{\bf x}) = C_k ({\bf x}) + D_k ({\bf x})
     \int^\eta \frac{d\eta^\prime}{a^2 Q},
    \label{e:unified-wave-sol-gen}
\eea
implying the conservation of $\Phi_k$ in the large-scale
limit. Since the coefficient $C_k({\bf x})$ does not depend
explicitly on $V(\phi)$, $f(\phi,R)$ or $\omega(\phi)$, the
perturbations are conserved independently of general changes in
the background equation of state and remain a priori conserved
even under changes of the underlying gravity. Since we haven't
specified a cosmological scenario yet, these results remain valid
for any four-dimensional FLRW background model which may be
derived from the general action eq.\ (\ref{e:general-action}). For
modes with physical sizes of order of the Hubble radius, however,
we stress that the conservation of the perturbed variable $\Phi_k$
may be delayed in a situation where the background evolution
becomes increasingly sensitive to the effects of the higher-order
corrections. Indeed, for both scalar and tensor metric
perturbations, the time-dependent variables $Q$ and $s$ may
eventually be altered by the presence of the curvature
corrections. In that case, the source $Q$ of the effective
potential for the perturbation may no longer grow monotonically
and may even decrease at the onset or during a high-curvature
regime. In the context of standard inflation models, a similar
effect has been associated in \cite{Leach} with an enhancement of
the amplitude of scalar metric perturbations just after the scales
cross the Hubble radius in some special situations. But it remains
from eq.\ (\ref{e:unified-wave-sol-gen}) that the evolution of the
perturbed variables should be frozen out for scales far beyond the
Hubble radius. Apparently, there is also no restriction on the
sign of the frequency shift occurring when the curvature
$(\alpha')$ corrections dominate the background dynamics: $s$ may
become negative or infinite depending on its particular form. One
interpretation of the denominator of the frequency shift (eq.\
(\ref{e:s_scalar}) and eq.\ (\ref{e:s_tensor}) for scalar and
tensor metric perturbations, respectively) is that as the
$\alpha'$ correction terms become comparable to the lowest-order
terms, it marks the breakdown of the approximation we have
adopted, in that we are entering a regime where we should include
all the higher-order corrections. Although we expect these
modifications to be dependent on the type of gravity chosen for
the background evolution, we now proceed to investigate further
the evolution of the perturbation variables inside the Hubble
radius, for we are interested in obtaining the spectrum as the
variables leave the Hubble radius during inflation.

\subsection{Quantum generation of vacuum fluctuations}
\label{ssec:quantum-generation}

In the semiclassical approximation \cite{Hwang:1994rz}, the
perturbed parts of the fields and metric are regarded as quantum
mechanical operators, whilst the background parts are considered
as classical. In the Heisenberg representation, where the quantum
operators (denoted with an overhat) carry the time-dependence, we
have for instance
\bea
   & & \phi({\bf x},t)=\bar\phi(t)+\delta\hat\phi({\bf x}, t),
       \quad
       \varphi({\bf x},t)\rightarrow\hat\varphi({\bf x},t),\quad
   \nonumber \\
   & & \hat \varphi_{\delta\phi} \equiv \hat \varphi -
       \frac{H}{\dot \phi}\delta \hat \phi,
       \quad
       \delta \hat \phi_\varphi \equiv \delta \hat \phi
       - {\dot \phi \over H} \hat \varphi,
       \quad {\rm etc.}
   \label{e:quantum-decomp}
\eea

Tensor metric perturbations, leading to the formation of a
cosmological stochastic background of gravitational waves, have
two polarisation states, whereas scalar metric perturbations have
a single component. For conciseness, we shall restrict the
description of the quantum generation to tensor-type
perturbations; being without the polarisation states the
scalar-type perturbation is simpler, and can be read from the
tensor-type analyses by ignoring the tensor and the polarisation
indices. It proves convenient to introduce a decomposition based
on the two polarisation states which, for a flat three-space
background, is given by \cite{Ford-Parker-1977,Hwang:1997uc}
\bea
   \hat \Phi_{ij} ({\bf x}, t)
   &\equiv& \int {d^3 k \over (2 \pi)^{3/2} }
       \hat \Phi_{ij} ({\bf x}, t; {\bf k})
    \\
   &\equiv&  \int {d^3 k \over (2 \pi)^{3/2} }
       \left[ \sum_\ell e^{i {\bf k} \cdot {\bf x}}
       \Phi_{\ell {\bf k}} (t)
       \hat a_{\ell {\bf k}} e^{(\ell)}_{ij} ({\bf k})
       + {\rm h.c.} \right].
   \label{e:mode-decomposition} \nonumber
\eea
Here, ${\rm h.c.}$ denotes the Hermitian conjugate, $\ell =
+,\times$ represent the two polarisation states and $\Phi_{\ell
{\bf k}}(t)$ is the mode function for the tensor metric
perturbation. The $e^{(\ell)}_{ij}({\bf k})$ form a base of
polarisation states which yields $e^{(\ell)}_{ij} ({\bf k})
e^{(\ell^\prime)\;ij} ({\bf k}) = 2 \delta_{\ell \ell^\prime}$.
The annihilation and creation operators $\hat a_{\ell {\bf k}}$
and $\hat a_{\ell {\bf k}}^\dagger$ of each polarisation state
satisfy the standard commutation relations on constant time
hypersurfaces 
\bea 
&& [ \hat a_{\ell {\bf k}},
   \hat a_{\ell^\prime {\bf k}^\prime} ]  = 0 =
   [ \hat a^\dagger_{\ell {\bf k}},
   \hat a^\dagger_{\ell^\prime {\bf k}^\prime} ],
   \nonumber \\
&& [ \hat a_{\ell {\bf k}},
   \hat a^\dagger_{\ell^\prime {\bf k}^\prime} ]
   = \delta_{\ell \ell^\prime} \delta^3 ({\bf k} - {\bf k}^\prime).
   \label{e:standard-commutation-relation}
\eea
By projecting $\hat \Phi_{ij} ({\bf x}, t)$ on the basis of
polarisation states, we easily obtain the mode expansion for each
polarisation state of the gravitational wave:
\bea
   \hat {\Phi}_\ell ({\bf x}, t)
&\equiv& {1 \over 2} \int {d^3 k \over (2 \pi)^{3/2} }
   \hat \Phi_{ij} ({\bf x}, t; {\bf k})
   e^{(\ell)\;ij} ({\bf k})
   \nonumber \\
&=&  \int {d^3 k \over (2 \pi)^{3/2} }
       \left[ e^{i {\bf k} \cdot {\bf x}} \Phi_{\ell {\bf k}} (t)
       \hat a_{\ell {\bf k}} + {\rm h.c.} \right].
   \label{e:mode-expansion}
\eea
The mode function $\Phi_{\ell {\bf k}} (t)$ is a complex
solution of the classical mode evolution equation. Replacing the
classical variable $\Phi$ with the Hilbert operator $\hat \Phi$ in
the second-order perturbed action eq.\ (\ref{e:action-2order})
leads to an equation for the mode function $\Phi_{\ell {\bf k}}$
which satisfies essentially the same form as the one obeyed by the
classical variable, namely eq.\ (\ref{e:unified-wave-cs}). From the
quadratic effective action in the perturbed variable, we then
derive the momentum $\hat \pi_{\Phi_\ell}$ canonically conjugate
to $\hat \Phi_\ell$, 
\bea
  \hat \pi_{\Phi_\ell} ({\bf x}, t) \equiv
  \frac{\partial {\cal L}}{\partial \dot {\hat \Phi}_\ell}
  = a^3 Q \dot {\hat\Phi}_\ell({\bf x}, t).
\eea
This implies that the equal-time commutation relation
$[\hat
\Phi_\ell({\bf x},t), \hat \pi_{\Phi_\ell} ({\bf x}^\prime, t)] =
i \delta^3 ({\bf x} - {\bf x}^\prime)$
leads to
\bea
   \left[  \hat \Phi_\ell ({\bf x},t),
   \dot {\hat \Phi}_\ell ({\bf x}^\prime, t) \right]
       = {i \over a^3 Q} \delta^3 ({\bf x} - {\bf x}^\prime).
   \label{e:commutation-relation}
\eea
Requiring agreement between
eq.\ (\ref{e:standard-commutation-relation}) and
eq.\ (\ref{e:commutation-relation}), the Wronskian of the mode
function $\Phi_{\ell {\bf k}} (t)$ must satisfy
\bea
{\cal W}_t \{\Phi_{\ell \bf k},\Phi_{\ell \bf k}^{*} \}\equiv
   \Phi_{\ell \bf k} \dot \Phi_{\ell \bf k}^{*}
   -\Phi^*_{\ell \bf k}\dot \Phi_{\ell \bf k} = \frac{i}{a^3 Q},
   \label{e:Wronskian}
\eea
where the index $t$ means that the derivatives in the
Wronskian are with respect to the cosmic time coordinate. The
power-spectrum based on the vacuum expectation value $(\hat
a_{\ell{\bf k}} \ket{0} \equiv 0,\,\forall\, {\bf k})$ is then
\bea 
   {\cal P}_{\hat \Phi_\ell} (k, t)
   &\equiv& {k^3 \over 2 \pi^2} \int
   \bra{0} \hat \Phi_\ell ({\bf x} + {\bf r}, t)
   \hat\Phi_\ell({\bf x},t)\ket{0} e^{-i{\bf k}\cdot{\bf r}}d^3 r
   \nonumber \\
   &=& {k^3 \over 2 \pi^2} |\Phi_{\ell k} (t)|^2.
   \label{e:power-spect-def}
\eea
In other words, the power-spectrum corresponds to the Fourier
transform of the two-point correlation function of the metric
fluctuations. In a space without any preferred direction, the two
polarisation states of the gravitational waves are expected to
contribute equally. Hence, we have
\bea
   {\cal P}_{\hat C_{ij}} ({\bf k}, t)
   = 2 \sum_\ell {\cal P}_{\hat C_\ell} ({\bf k}, t) = 2
   \sum_\ell {k^3 \over 2 \pi^2} \left|\hat C_{\ell {\bf k}}
   (t)\right|^2.
   \label{P-tensor}
\eea
The power-spectrum of scalar-metric
perturbations can be derived in a similar manner, although no
polarisation decomposition is required \cite{GGT-quantum}.
We find
\bea
   {\cal P}_{\hat\varphi_{\delta\phi}} ({\bf k}, t)
   = {k^3 \over 2 \pi^2} \left|\hat\varphi_{\delta\phi\,
   {\bf k}} (t)\right|^2.
   \label{P-scalar}
\eea
We now turn our attention to the spectral distributions $|\hat C_{\ell
{\bf k}} (t)|$ and $|\hat\varphi_{\delta\phi\, {\bf k}} (t)|$.

Depending on the sign of the frequency shift $s=\pm \tilde s$
(which we shall consider constant in a first approximation),
eq.\ (\ref{e:unified-Bessel-wave}) becomes either a Bessel
equation (for positive $s$) or a modified Bessel equation (for
negative $s$). For positive $s$, the normalised mode function
solution of eq.\ (\ref{e:unified-Bessel-wave}) corresponds to a
superposition of Hankel functions of the first and second kind,
\bea
   \Phi_k = {\sqrt{\pi} \over 2} {\sqrt{|\eta|} \over a \sqrt{Q}}
       \left[ c_1 (k) H_\nu^{(1)} (x)
       + c_2 (k) H_\nu^{(2)} (x) \right],
   \label{positive-sol}
\eea
with $x\equiv\sqrt{\tilde s} k |\eta|$ from which the usual
condition between the coefficients $|c_2|^2 - |c_1|^2 = 1$
is obtained from eq.\ (\ref{e:Wronskian}). In the large-scale limit,
$\sqrt{\tilde s} k |\eta| \ll 1$, we have, for $\nu \neq 0$ and
$\nu=0$ respectively,
\bea
   & & {\cal P}_{\hat \Phi}^{1/2}
       = {1 \over \sqrt{Q}} {H \over 2 \pi}
       {{\tilde s}^{-\nu/2} \over a H |\eta|}
       {\Gamma(\nu) \over \Gamma(3/2)}
       \left( {k |\eta| \over 2} \right)^{3/2 - \nu},
   \label{P-1} \\
   & & {\cal P}_{\hat \Phi}^{1/2}
       = {2 \sqrt{|\eta|} \over a \sqrt{Q}}
       \left( {k \over 2 \pi} \right)^{3/2}
       \ln( \sqrt{|s|} k |\eta|).
   \label{P-2}
\eea
Here we have considered the conventional choice of vacuum
$c_2 = 1$ and $c_1 = 0$, corresponding to positive frequency in
the asymptotic flat space time for $\eta \rightarrow -\infty$. An
additional $\sqrt{2}$ factor appears for the gravitational wave
power spectrum ${\cal P}_{\hat C_{ij}}^{1/2}$ which follows from
proper consideration of the two polarisation states; see eqs.
(\ref{P-tensor},\ref{P-scalar}).

{}For the case of negative $s$, we have the modified Bessel functions
$I_\nu (x)$ and $K_\nu (x)$ as the two independent solutions of
eq.\ (\ref{e:unified-Bessel-wave}), that is
\bea
   \Phi_k = {\sqrt{|\eta|} \over a \sqrt{Q}}
       \Bigl[ c_1 (k) I_\nu (x)
       + c_2 (k) K_\nu (x) \Bigr],
   \label{negative-sol}
\eea
with $x\equiv\sqrt{\tilde s} k |\eta|$ and an unusual
condition upon the coefficients $c_1^* c_2 - c_1 c_2^* = i$ arising
from eq.\ (\ref{e:Wronskian}). In the large scale limit, the power
spectrum exhibits similar dependence on the wave number as in
eqs.\ (\ref{P-1})--(\ref{P-2}). However, modes evolving inside the Hubble
radius face an instability since
\bea
   I_\nu (x) \sim {1 \over \sqrt{2 \pi x}} e^x, \quad
   K_\nu (x) \sim \sqrt{\pi \over 2 x} e^{-x},
\eea
which may lead to invalidate our assumption of small
perturbations, hence the use of linear approach to cosmological
perturbations. This is the condition we alluded to earlier, as
marking a breakdown in the underlying action, requiring us to
replace it with even higher-order corrections to enable us to deal
properly with a high curvature epoch. Fortunately, a number of
features emerge in the weak coupling regime which allows us to
proceed further.

\section{An example: the pre-Big Bang scenario}
\label{sec:PBB}

The observationally relevant scales are thought to have exited the
Hubble radius within about $60$ $e$-folds before the end of
inflation. We have in mind a scenario where inflation occurs in
our generalised gravity. This is then followed by a smooth
transition to the ordinary radiation dominated era of the standard
FLRW model. In the large scale limit, the perturbation
variable $\Phi$ is conserved independently of changes in the
underlying gravity theory, and the power-spectra based on the quantum
vacuum expectation value can be identified at a later epoch as the
classical power-spectra based on the classical volume average.
Thus, eqs.\ (\ref{P-1})--(\ref{P-2}) are now
considered valid for the classical power-spectra and the spectral
index of the scalar and tensor-type perturbations become
\cite{GGT-quantum} 
\bea
   n_S - 1 = 3 - 2\nu_s,   \quad
   n_T= 3 - 2\nu_t.        \label{n-spectra}
\eea
Here, $\nu_s$ and $\nu_t$ are to be evaluated at the time of
Hubble radius crossing during inflation, and will thus directly
reflect the kinematics of the background when the perturbations
exit the Hubble radius.

\begin{figure}[ht]
   \centering
   \leavevmode
   \epsfysize=5cm
   \epsfbox{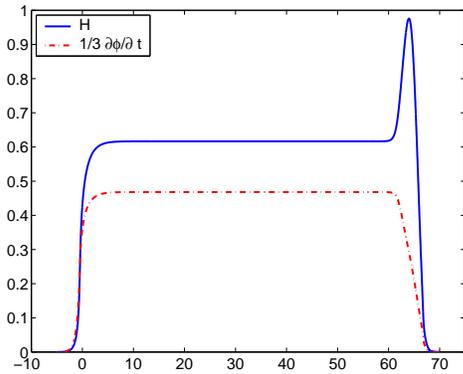}\\
   \caption[Non-singular evolutions based on high-order corrections]
   {\label{f:non-sing}
Here we reproduce (from \cite{Cartier:2001gc}) a non-singular
evolution for the Hubble parameter $H=\dot{a}/a$ and for
$\dot{\phi}/3$, as a function of the number of e-folds, $N=\ln a$.
The low-energy, dilaton-driven phase takes place approximately for
$-\infty < N \leq -3$. After a short transition, this initial
period is followed  by a string phase with nearly constant Hubble
parameter and linearly growing  dilaton, for $2 \leq N \leq 55$.
After a successful exit triggered by loop corrections, the
background evolution enters the FLRW radiation-dominated phase at
$N \simeq 68$.
    }
\end{figure}

As an application of our general results, we may consider in
particular the pre-Big Bang scenario of string cosmology
\cite{PBB}. In our formalism, it corresponds to the case $f =
e^{-\phi} R$, $\xi = e^{-\phi}= -\omega$ and $V = 0$. Examples of
regular backgrounds have recently been obtained by
supplementing the low-energy effective action of string theory by
the kind of higher-order corrections given in eq.\
(\ref{e:action_corr}) \cite{pbb-higher-corr,Cartier:1999vk}. Fig.\
\ref{f:non-sing} is a typical example of non-singular evolution:
when the curvature scale is of order of the inverse string scale,
the $\alpha'$ corrections may eventually stabilise the background
into a de-Sitter like regime of constant Hubble parameter $H\sim
{\cal O}(\lambda_s^{-2})$ and linearly growing  (in cosmic time)
dilaton. Then, quantum loop corrections (based on the string
coupling expansion) or a non-perturbative potential (yet to be
determined) may trigger an exit to the FLRW radiation-dominated
phase, with constant dilaton field.

It has been known for some time that this cosmological model
derived from string theory predicts a very blue spectra for both
the scalar- and tensor-type perturbations
\cite{Brustein:1995kn,Hwang:1998fp,Copeland:1998ie,Lidsey:1999mc}.
Recalling the (conformal time) expression of the tree-level
solutions \cite{Gasperini:1994xg},
\bea a(\eta) =
(-\eta)^{\frac{1}{2}(1-\sqrt{3})},    \quad \phi(\eta) = -\sqrt{3}
\ln (-\eta), \label{e:low_sol}
\eea
the exponents of the scalar and tensor pump fields coincide,
$\gamma_s=\gamma_t = 1/2$, hence $\nu_s=\nu_t=0$.
Using eq.\ (\ref{n-spectra}), we then obtain
\bea
   n_S - 1 = n_T = 3,
\eea
whereas the observationally favoured Harrison-Zel'dovich type
spectra correspond to $n_S \simeq 1$ and $n_T \simeq 0$. Hence,
the pole-like acceleration expansion (in the string frame) of the
pre-Big Bang scenario cannot, in principle, be considered as the
source for the observed fluctuations in the CMBR and also the
observed large scale structures. A possible resolution of this
problem for the scalar perturbations was proposed in
\cite{Copeland:1997ug}, where fluctuations of the axion field
present in the low-energy string action can generate the observed
spectral index. There also remains a possibility that the
high-curvature regime can last long enough so that all the
relevant scales that we observe today actually exit the Hubble
radius during this era.

To determine whether a high-curvature regime may or may not
flatten the spectral slope, we have to follow the evolution of the
large frequency modes, associated with small physical wavelengths.
We start by recalling that during the high-curvature regime, both
the Hubble rate and the dilaton field become time-independent,
$H\equiv \tilde H$ and $\dot\phi \equiv \tilde \phi$. Hence, for
both scalar- and tensor-type perturbations, the potential reduces
to its tree-level (TL) form and we can express the pump field in
the convenient form, $z= const \times z_{TL}$. Leaving aside the
frequency shift, i.e. setting $s\equiv1$ in eq.\
(\ref{e:unified-wave-cf}), it is already well known
\cite{Brustein:1995ah} that the spectral slope of the
high-frequency modes, crossing the Hubble radius in the high
curvature, stringy regime, and re-entering in the radiation era,
is fully determined by the fixed point values $\tilde \phi, \tilde
H$. Indeed, during such a string phase, the scale factor undergoes
the usual de Sitter exponential expansion, while the logarithmic
evolution of the dilaton, in conformal time, is weighted by the
ratio $\tilde \phi/\tilde H$, i.e. $\phi(\eta) \sim - (\tilde \phi
/\tilde H) \log (-\eta)+$const \cite{Gasperini:1995fm}. By
introducing the convenient shifted variable $\tilde
{\overline{\phi}} \equiv \tilde {\phi} - 3 \tilde H$ and the ratio
$\vartheta \equiv \tilde {\overline{\phi}}/\tilde H$, and
referring the spectrum to a fixed point allowing a subsequent
(loop catalysed) exit \cite{Cartier:1999vk}, i.e. $-1\leq
\vartheta \leq 0$, one easily finds that the exponents of the pump
fields satisfy $0 \leq \gamma_s = \gamma_t =
\frac{1}{2}\left(1+\vartheta \right)\leq 1/2$. Hence, the spectral
indices for the high-frequency modes, leaving the Hubble radius
during the de-Sitter like era and re-entering in the radiation
dominance epoch is given by
\bea
  2 \leq  n_S - 1 = n_T \leq 3.     \label{e:indices}
\eea
Although still not satisfying the observational bounds, they provide
an indication of a possible resolution for suitable higher-order
corrections.

\begin{figure}[t]
   \centering
   \leavevmode
   \epsfysize=5cm
   \epsfbox{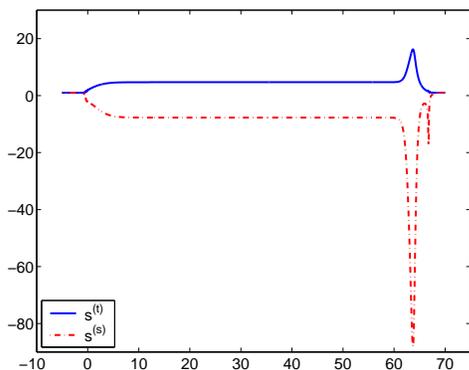}\\
   \caption[Frequency shifts $s$ for scalar and tensor-type perturbations]
   {\label{f:shift}
Here we compare the frequency shifts for scalar and tensor-type
perturbations, $s^{(s)}$ and $s^{(t)}$ respectively, in the
regular background of Fig.\ \ref{f:non-sing}. The frequency shift
for tensor-type perturbations is found to be always positive,
whereas $\alpha'$ corrections are responsible for the sign change
of the frequency shift of scalar-type perturbations. The latter
may imply an exponential instability for the perturbation
amplitude of the comoving mode whose physical wavelength is
smaller than the size of the Hubble radius during the stringy
high-curvature regime.
    }
\end{figure}

In Fig.\ \ref{f:shift}, we illustrate the evolution of the
frequency shifts given by eq.\ (\ref{e:s_scalar}) and eq.\
(\ref{e:s_tensor}) in the regular background of Fig.\
\ref{f:non-sing}. In the asymptotic past, the curvature
corrections are negligible, i.e. $\alpha^\prime \rightarrow 0$ and we
have $s^{(s)} = s^{(t)} = 1$. At the onset of the high-curvature
regime, the behaviour of the frequency shift for scalar and
tensor-type perturbations is strongly affected by the $\alpha'$
corrections, $s^{(t)}$ is attracted to a positive value
while $s^{(s)}$ decreases and becomes even negative. Then, when
the universe reaches the FLRW radiation-dominated era with a
constant dilaton field, the higher-order corrections identically
vanish and we recover $s^{(s)} = s^{(t)} = 1$.

{}For tensor-type perturbations, the influence of the (all-time
positive) frequency shift has been recently investigated in
\cite{Cartier:2001gc}. Its contribution has been found to amount
to an overall rescaling, by a numerical factor of order unity, of
the total energy density of the background, but does not alter the
spectral index of the high-frequency modes in eq.\
(\ref{e:indices}). For scalar-type perturbation, however, the
situation is more complex, due to the sign change of the effective
frequency shift during the high-curvature regime. For a comoving
mode evolving inside the Hubble radius at the onset of the
high-curvature phase, the sign change is a priori responsible for
turning the oscillatory behaviour of the perturbation variable
into exponential growth or exponential suppression. In the latter
case, this may imply an abrupt end-point in the spectral
distribution, where the highest frequency mode to get amplified
corresponds to the one which left the Hubble radius just before
the onset of the string high-curvature phase. In the former case,
on the contrary, the growth of the amplitude of the perturbation
may be exponential until it exits the Hubble radius. Neither case
seems satisfactory, and the change of sign is really an indication
that the scalar fluctuations should be considered in the light of
the full string action in this regime. For example, including
higher-order curvature contributions to probe the highly-curved
regime, we have focused on the next to leading order in the
infinite series of the inverse string tension expansion. Although
this represents a significant improvement compared to previous
studies, it still places a limit to our ability to analyse the
results. It is natural to ask how sensitive our results are with
respect to the inclusion of even higher-order derivative terms,
such as $R^3$ or $R_{\mu\nu;\rho}R^{\mu\nu;\rho}$. To estimate the
impact of such terms, while conserving equations of motion to
second-order in the fields, we have looked at the relative
contribution of $(\partial_\mu \phi)^4$ and $(\partial_\mu
\phi)^6$. Indeed this very preliminary analysis indicates that the
frequency shift may no longer be negative by including the
infinite series expansion. The tensor contributions are not
affected by this kind of higher order terms.

\section{Conclusion}
\label{sec:Conclusion}

In this paper, we have for the first time presented a complete set
of equations describing the classical evolution of scalar, vector
and tensor perturbations in the context of a non-singular
cosmology arising out of higher-order corrections to the
low-energy string action. From these we have obtained the
large-scale exact solutions for both the scalar and
tensor perturbations, the associated large-scale conserved
quantities and shown how there is general conservation of the
angular momentum for the vector perturbations in the absence of
dissipative  processes. The perturbed actions for the scalar- and
tensor-type perturbations have been obtained, thus laying ground
work for future applications to the quantum generation processes.
In particular we have calculated the final spectra of the scalar-
and tensor-type perturbations generated from the vacuum quantum
fluctuations of the metric and the field, given the non-singular
background evolution during the kinetic driven inflation.

Our results are valid for the general action of eq.\
(\ref{e:general-action}), and for the general perturbed FLRW
metric in eq.\ (\ref{e:perturbed-metric}). Moreover our results up
to \S \ref{sec:QuantumGen} describing the classical evolution of
the background and the perturbations are generally valid without
assuming any cosmological scenario, like the pre-Big Bang or other
types of inflation models. In order to calculate the initial
conditions for the seed structures generated from the quantum
fluctuations, we need to specify the specific cosmological
scenario for the background. As an example, we have applied our
general results to the case of the pre-Big Bang scenario of string
cosmology. Our results confirm previous expectations, that the low
frequency modes, crossing the Hubble radius in the low-curvature
regime, are unaffected by higher-order corrections. However, the
higher-order corrections generally reduces the spectral indices
for the high-frequency modes which are leaving the Hubble radius
during the de-Sitter like era. Although still not satisfying the
observational bounds, they provide an indication of a possible
resolution for suitable higher-order corrections.

\section*{Acknowledgments}

It is a pleasure to thank M.~Gasperini and G.~Veneziano for
helpful and stimulating discussions. CC is supported by the Swiss
NSF, grant No. 83EU-054774 and ORS/1999041014. JH is supported by
the Korea Research Foundation Grants (KRF-2000-013-DA004 and
2000-015-DP0080).



\begin{references}
\bibitem{CMBR}
         P.~de Bernardis et al., Nature {\bf 404}, 955 (2000);
         A.~Balbi et al., Astrophys. J. {\bf 545}, L1 (2000);
         A.T.~Lee et al., astro-ph/0104459;
         C.B.~Netterfield et al., astro-ph/0104460.
\bibitem{Inflation}
         E.W.~Kolb and M.S.~Turner, {\it The Early Universe},
         (Addison-Wesley, Redwood City, CA, 1990);
         A.D.~Linde, {\it Particle Physics and Inflationary Cosmology},
         (Harwood, New York, 1990).
\bibitem{I-grav}
         M.D.~Pollock and D.~Sahdev, Phys. Lett. B {\bf 222}, 12 (1989).
\bibitem{SC-grav}
         J.J.~Levin and K.~Freese, Nucl. Phys. B {\bf 421}, 635 (1994).
\bibitem{PBB}
         G.~Veneziano, Phys. Lett. B {\bf 265}, 287 (1991);
         M.~Gasperini and G.~Veneziano, Astroparticle Phys.
         {\bf 1}, 317 (1993).
         An updated collection of papers and references on the
         pre-Big Bang scenario is available at the URL
         ``http://www.to.infn.it/$\sim$gasperin/''.
\bibitem{Brustein:1995kn}
         R.~Brustein et al., Phys. Rev. D {\bf 51}, 6744 (1995).
\bibitem{Hwang:1998fp}
         J.~Hwang, Astropart. Phys. {\bf 8}, 201 (1998).
\bibitem{Copeland:1998ie}
         E.J.~Copeland, J.E.~Lidsey and D.~Wands,
         Phys. Lett. B {\bf 443}, 97 (1998).
\bibitem{Lidsey:1999mc}
         J.~Lidsey, D.~Wands and E.J.~Copeland,
         Phys. Rept. {\bf 337}, 343 (2000).
\bibitem{Copeland:1997ug} E.J.~Copeland, R.~Easther and D.~Wands,
         Phys. Rev. D {\bf 56}, 874 (1997).
\bibitem{backreaction}
         S.-J.~Rey, Phys. Rev. Lett. {\bf 77}, 1929 (1996);
         A.~Ghosh, R.~Madden and G.~Veneziano,
         Nucl. Phys. B {\bf 570}, 207 (2000).
\bibitem{pbb-higher-corr}
         S.~Foffa, M.~Maggiore and R.~Sturani,
         Nucl. Phys. B {\bf 552}, 395 (1999);
         R.~Brustein and R.~Madden, JHEP {\bf 07}, 006 (1999).
\bibitem{Cartier:1999vk}
         C.~Cartier, E.J.~Copeland and R.~Madden,
         JHEP {\bf 01}, 035 (2000).
\bibitem{Maggiore:1997bg}
         M.~Maggiore and R.~Sturani, gr-qc/9705082.
\bibitem{Kawai}
         S.~Kawai, M.~Sakagami and J.~Soda, gr-qc/9901065;
         S.~Kawai and J.~Soda, Phys. Lett. B {\bf 460}, 41 (1999).
\bibitem{Gasperini:1997up}
         M.~Gasperini, Phys. Rev. D {\bf 56}, 4815 (1997).
\bibitem{Hwang:1999gf}
         J.~Hwang and H.~Noh, Phys. Rev. D {\bf 61}, 043511 (2000).
\bibitem{Choi:1999zy}
         K.~Choi, J.~Hwang and K.W.~Hwang,
         Phys. Rev. D {\bf 61}, 084026 (2000).
\bibitem{Hwang:1991aj-2001}
         J.~Hwang, Astrophys. J. {\bf 375}, 443 (1991);
         J.~Hwang and H.~Noh, astro-ph/0102005.
\bibitem{Metsaev:1987zx}
         R.~Metsaev and A.~Tseytlin,
         Nucl. Phys. B {\bf 293}, 385 (1987).
\bibitem{Field_DeWitt}
         B.~DeWitt,  {\it Dynamical Theory of Groups and Fields},
         (Blackie \& Son Ltd., London, 1965).
\bibitem{covariant-T}
         J.~Ehlers, Gen. Rel. Grav. {\bf 25}, 1225 (1993);
         G.F.R.~Ellis in {\it General Relativity and Cosmology},
         ed. R. K. Sachs (New York: Academic), p. 104 (1971);
         in {\it Cargese Lectures in Physics}, ed. E. Schatzmann
         (New York: Gordon and Breach), p. 1 (1973).
\bibitem{Pert_Bardeen}
         J.M.~Bardeen, in {\it Particle Physics and Cosmology},
         (Gordon and Breach, London, 1988).
\bibitem{Hwang-MSF}
         J.~Hwang, Astrophys. J. {\bf 427}, 542 (1994);
         J.~Hwang and H.~Noh, Phys. Rev. D {\bf 54}, 1460 (1996).
\bibitem{Lukash-1980}
         V.N.~Lukash, Sov. Phys. JETP Lett. {\bf 31}, 596 (1980);
         Sov. Phys. JETP {\bf 52}, 807 (1980).
\bibitem{Mukhanov-1988}
         V.F.~Mukhanov, Sov. Phys. JETP {\bf 68}, 1297 (1988).
\bibitem{Hwang-Noh-2001-fR}
         J.~Hwang, J. Korean Phys. Soc., {\bf 35}, S633 (1999),
            astro-ph/9909150;
         J.~Hwang and H.~Noh, Phys. Lett. B {\bf 506}, 13 (2001).
\bibitem{Bardeen:1980kt}
         J.M.~Bardeen, Phys. Rev. D {\bf 22}, 1882 (1980).
\bibitem{Cartier:2001gc}
         C.~Cartier, E.J.~Copeland and M.~Gasperini, gr-qc/0101019.
\bibitem{Grishchuk:1975ny}
         L.P.~Grishchuk, Sov. Phys. JETP {\bf 40}, 409 (1975).
\bibitem{Hwang:1993cv}
         J.~Hwang, Phys. Rev. D {\bf 48}, 3544 (1993).
\bibitem{Grishchuk:1990bj-1992tw}
         L.P.~Grishchuk and Y.V.~Sidorov,
         Phys. Rev. D {\bf 42}, 3413 (1990);
         L.P.~Grishchuk, H.A.~Haus and K.~Bergman,
         Phys. Rev. D {\bf 46}, 1440 (1992).
\bibitem{Leach}
         S.~Leach and A.~Liddle, Phys. Rev. D {\bf 63}, 043508 (2001);
         S.~Leach, M.~Sasaki, D.~Wands and A.~Liddle, astro-ph/0101406.
\bibitem{Hwang:1994rz}
         J.~Hwang, Class. Quantum Grav. {\bf 11}, 2305 (1994).
\bibitem{Ford-Parker-1977}
         L.H.~Ford and L.~Parker, Phys. Rev. D {\bf 16}, 1601 (1977).
\bibitem{Hwang:1997uc}
         J.~Hwang, Class. Quantum Grav. {\bf 15}, 1401 (1998).
\bibitem{GGT-quantum}
         J.~Hwang, Class. Quantum Grav. {\bf 14}, 3327 (1997);
         J.~Hwang and H.~Noh, Class. Quant. Grav. {\bf 15}, 1387 (1998).
\bibitem{Gasperini:1994xg}
         M.~Gasperini and G.~Veneziano,
         Phys. Rev. D {\bf 50}, 2519 (1994).
\bibitem{Brustein:1995ah}
        R.~Brustein et al., Phys. Lett. B {\bf 361}, 45 (1995).
\bibitem{Gasperini:1995fm}
        M.~Gasperini, M.~Giovannini and G.~Veneziano,
        Phys. Rev. D {\bf 52}, 6651 (1995).
\end{references}
\end{document}